\documentclass[11pt]{article}
\usepackage{amsmath}
\usepackage{graphicx}
\begin{document}
\title{Normal Mode Expansion of Damped Coupled Oscillators in 3-Dimensions}
\author{Milind Shyani \\ Department of Physics \\ BITS Pilani KK Birla Goa Campus\\ \texttt{f2009675@bits-goa.ac.in}}
\date{\today}
\maketitle
\vspace{0.05cm}
\begin{abstract}
In this paper, I aim to study free oscillations of a system of oscillators in more than one dimensions in the absence of damping. The basic approach lies in decoupling the motion in the individual perpendicular directions. Once the equations are decoupled, the existent techniques of Normal mode expansion for 1-dimensional oscillators are used to solve for the equations of motion. I also study the motion of a driven system of oscillators in higher dimensions in the presence of a velocity dependent damping force.
\end{abstract}
\vspace{0.25cm}
\section{Introduction}
Motion of a driven 1-dimensional system of N oscillators in the presence of damping has been well analyzed in the framework of Newtonian mechanics. However, it has not had the same attention in higher dimensions. The basic problem lies in the fact that since the motion of an oscillator in a lattice is influenced by its nearest neighbors, the equations of motion are coupled amongst the perpendicular directions, viz., $x$, $y$ and $z$ components. However under the approximation that the perturbations are small, the motion can be decoupled in the individual directions. I discuss the problem of free oscillations in the first section and introduce damping and driving terms in the following section.
\section{Undamped Natural Oscillations in 2D}
In this section, I will discuss the effect of a perturbation on a 2-dimensional lattice of bodies identical in mass. The lattice under consideration is a finite $(N\times{M})$ structure. I deal with systems under natural oscillations, viz., under the absence of damping and external forces. Every oscillator is connected with its closest neighbor with a linear spring. For simplicity, the lattice is understood to lie in the first quadrant of the regular cartesian co-ordinate system. For integral values of $i$ and $j$, the ordered pair $(i,j)$ denotes a body belonging to the $i^{th}$ column and $j^{th}$ row. The initial distance between any two closest neighbors is equal to the unstretched length of the spring, $a$. So the position vector $\vec{r}_{i,j}$ of an oscillator $(i,j)$ can be written as
\begin{align}
\vec{r}_{i,j} &= i\cdot{a}\hat{x} + j\cdot{a}\hat{y} \nonumber
\end{align}
On a small perturbation, the new position vector of an oscillator $(i,j)$, $\vec{r_{i,j}}$ is given by 
\begin{align}
\vec{r}_{i,j} &= i\cdot{a} \hat{x} + j\cdot{a} \hat{y} + \vec{r'}_{i,j}
\end{align}
The equation of motion for a body not on any of the boundaries can be written as 
\begin{align}
m\frac{d^2\vec{r}_{i,j}}{dt^2} &= \vec{F}_{(i-1,j),(i,j)} + \vec{F}_{(i+1,j),(i,j)} + \vec{F}_{(i,j-1),(i,j)} + \vec{F}_{(i,j+1),(i,j)}
\end{align}
where ${\omega_{0}}^2$ is the spring constant and m is the mass of the oscillator. Note that the convention used here is that of Goldstein; $F_{\mu\nu}$ represents the force on the particle $\nu$ by the particle $\mu$. $F_{(i-1,j),(i,j)}$ can thus be expressed as
\begin{align}
\vec{F}_{(i-1,j),(i,j)} = -{{\omega_{0}}^2}\big[\mid \vec{r}_{i,j} - \vec{r}_{i-1,j}\mid - a\big]\frac{\vec{r}_{i,j} - \vec{r}_{i-1,j}}{\mid\vec{r}_{i,j} - \vec{r}_{i-1,j}\mid}
\end{align}
For small perturbations
\begin{align}
\frac{\vec{r}_{i,j} - \vec{r}_{i-1,j}}{\mid\vec{r}_{i,j} - \vec{r}_{i-1,j}\mid}
&\approx {\frac{\vec{r}_{i,j} - \vec{r}_{i-1,j}}{a}} \nonumber
\end{align}
Using (1),
\begin{align}
\mid\vec{r}_{i,j} - \vec{r}_{i-1,j}\mid &= \sqrt{(x_{i,j} - x_{i-1,j})^2 + (y_{i,j}- y_{i-1,j})^2} \nonumber \\
&= \sqrt{(a+x'_{i,j} - x'_{i-1,j})^2 + (y'_{i,j}- y'_{i-1,j})^2} \nonumber
\end{align}
For small perturbations, $O(x'^2/a^2)\approx{0}$
\begin{align}
\mid\vec{r}_{i,j} - \vec{r}_{i-1,j}\mid = a\sqrt{1 + \frac{2x'_{i,j}}{a} - \frac{2x'_{i-1,j}}{a}} \nonumber
\end{align}
On binomial expansion of the square-root term under the above approximation and using (3), force can be written as
\begin{align}
\mid\vec{r}_{i,j} - \vec{r}_{i-1,j}\mid - a = a\big[{1 + \frac{x'_{i,j}}{a} - \frac{x'_{i-1,j}}{a}}\big]-a \nonumber \\
\vec{F}_{(i-1,j),(i,j)} = -{{\omega_{0}}^2}[x'_{i,j} - {x'_{i-1,j}}\big]\frac{\vec{r}_{i,j} - \vec{r}_{i-1,j}}{a}
\end{align}
Expressing the position vector in terms of its components and using (1),
\begin{align}
\frac{\vec{r}_{i,j} - \vec{r}_{i-1,j}}{a} &= {(1+\frac{{x'_{i,j}}}{a} - \frac{x'_{i-1,j}}{a})\hat{x} + {(\frac{y'_{i,j}}{a}- \frac{y'_{i-1,j}}{a})}\hat{y}} \nonumber
\end{align}
Neglecting terms of order $O(x'^2/a^2)$, $O(y'^2/a^2)$ and $O(x'y'/a^2)$, (4) can be expressed as
\begin{align}
\vec{F}_{(i-1,j),(i,j)} = -{\omega^2_{0}}[x'_{i,j} - x'_{i-1,j}] \nonumber
\end{align}
The other terms in (2) can be evaluated similarly and the equation of motion for $(i,j)$ can be written as
\begin{align}
m\frac{d^2\vec{r}_{i,j}}{dt^2} = -{\omega^2_{0}}\big[(2x'_{i,j} - x'_{i-1,j} - x'_{i+1,j})\hat{x} + (2y'_{i,j} - y'_{i,j-1} - y'_{i,j+1})\hat{y}\big] \nonumber
\end{align}
One can see that the equation of motion is decoupled into two individual motions, about $\hat{x}$ and $\hat{y}$ such that,
\begin{align}
m\frac{d^2{x}_{i,j}}{dt^2} \equiv m\frac{d^2{x'}_{i,j}}{dt^2} = -{\omega^2_{0}}[2x'_{i,j} - x'_{i-1,j} - x'_{i+1,j}]\\
m\frac{d^2{y}_{i,j}}{dt^2} \equiv m\frac{d^2{y'}_{i,j}}{dt^2} = -{\omega^2_{0}}[2y'_{i,j} - y'_{i-1,j} - y'_{i+1,j}]
\end{align}
Once the motion is decoupled in the individual directions, the problem reduces to the familiar system of N coupled oscillators in one dimension \cite{1}. Assuming a test solution of the form,
\begin{align}
x'_{i,j} = A_{i,j}\cos(\omega t)
\end{align}
and plugging it in the equation of motion (5) one gets
\begin{align}
\frac{A_{i,j}}{A_{i+1,j} + A_{i-1,j}} = \frac{\omega^2_{0}}{-\omega^2 + 2\omega^2_{0}}
\end{align}
Since the ratio of the amplitudes is constant, one can assume ${A_{i,j}}$ to be of the form,
\begin{align}
{A_{i,j}} = C\sin(i\theta)\cos(j\phi)
\end{align}
On substituting the above expression in (8),
\begin{align}
\omega= 2\omega_{0}\sin\bigg(\frac{\theta}{2}\bigg)
\end{align}
The values of $\theta$ and $\phi$ can be found by imposing the boundary condition that the ends of the lattice are held fixed, hence
\begin{align}
x_{0,j} = 0    \hspace{0.5cm}\&\hspace{0.5cm} x_{M-1,j} = 0 \nonumber \\
x_{i,0} = 0    \hspace{0.5cm}\&\hspace{0.5cm} x_{i,N-1} = 0  \nonumber
\end{align}
On considering the boundary values, it is seen that $\theta$ and $\phi$ need to obey
\begin{center}
$\sin\big((M-1)\theta\big)=0$ $\hspace{0.5cm} \& \hspace{0.5cm} \sin\big((N-1)\theta\big)=0$
\end{center}
Hence, 
\begin{align}
\theta &= \frac{m\pi}{M-1} \hspace{0.5cm} where~m=0,1,\ldots,M-1\\
\phi &= \frac{n\pi}{N-1}  \hspace{0.5cm} where~n=0,1,\ldots,N-1
\end{align}
The indices, $m$ and $n$ correspond to the particular normal mode of the system. One can see that for a $N\times{M}$ lattice of oscillators, there exists $MN$ normal modes. Thus the motion of the system can be described by(up to an arbitrary phase factor)
\begin{align}
x' = \sum_{i=0}^{M-1} \sum_{j=0}^{N-1}C_{mn}\sin\bigg(i\frac{m\pi}{M-1}\bigg)\sin\bigg(j\frac{n\pi}{N-1}\bigg)\cos(\omega_{m}t)
\end{align}
$\omega$ carries an index m, since by $(11)$ the value of $\theta$ for different values of $m$ are different. So by $(10)$ the frequency is seen to depend on $m$. One can similarly treat the motion in $y$ with the boundary conditions that the ends are held fixed and obtain(up to an arbitrary phase factor),
\begin{align}
\omega_{n} &= 2\omega_{0}\sin{\bigg(\frac{\phi}{2}\bigg)}
\end{align}
\begin{align}
y' &= \sum_{i=0}^{M-1} \sum_{j=0}^{N-1}D_{mn}\sin\bigg(i\frac{m\pi}{M-1}\bigg)\sin\bigg(j\frac{n\pi}{N-1}\bigg)\cos(\omega_{n}t)
\end{align}
In the above expression $\theta$ and $\phi$ take the same values as (11) and (12). The values of $C_{mn}$ and $D_{mn}$ can be found by imposing initial conditions.
\subsection{Undamped Natural oscillations in 3D}
Natural oscillations in 3-Dimensions can be treated in a similar way. For a $M\times{N}\times{P}$ lattice under the boundary conditions that the ends are held fixed, following the convention developed in the parent section(up to an arbitrary phase factor),
\begin{align}
x' = \sum_{i=0}^{M-1} \sum_{j=0}^{N-1}\sum_{k=0}^{P-1}C_{mnp}\sin\bigg(i\frac{m\pi}{M-1}\bigg)\sin\bigg(j\frac{n\pi}{N-1}\bigg)\sin\bigg(k\frac{p\pi}{P-1}\bigg)\cos(\omega_{m}t)\nonumber\\
y' = \sum_{i=0}^{M-1}\sum_{j=0}^{N-1}\sum_{k=0}^{P-1}D_{mnp}\sin\bigg(i\frac{m\pi}{M-1}\bigg)\sin\bigg(j\frac{n\pi}{N-1}\bigg)\sin\bigg(k\frac{p\pi}{P-1}\bigg)\cos(\omega_{n}t)\nonumber\\
z' = \sum_{i=0}^{M-1}\sum_{j=0}^{N-1}\sum_{k=0}^{P-1}F_{mnp}\sin\bigg(i\frac{m\pi}{M-1}\bigg)\sin\bigg(j\frac{n\pi}{N-1}\bigg)\sin\bigg(k\frac{p\pi}{P-1}\bigg)\cos(\omega_{p}t)\nonumber
\end{align}
The frequencies are given as
\begin{align}
\omega_m= 2\omega_{0}\sin(\frac{\theta}{2}) \hspace {0.4cm}
\omega_n= 2\omega_{0}\sin(\frac{\phi}{2}) \hspace {0.4cm}
\omega_p= 2\omega_{0}\sin(\frac{\psi}{2})
\end{align}
where
\begin{align}
\theta &= \frac{m\pi}{M-1}\hspace{0.5cm} where~m=0,1,\ldots,M-1\\
\phi &= \frac{n\pi}{N-1}\hspace{0.5cm} where~n=0,1,\ldots,N-1\\
\psi &= \frac{p\pi}{P-1}\hspace{0.5cm} where~p=0,1,\ldots,P-1
\end{align}
\section{Forced oscillations with Damping}
In this section, I aim to study the motion of a system in the presence of a first order velocity dependent damping force and an external periodic force. I shall once again consider a $M\times{N}$ lattice as discussed in the previous section. If one tries to apply the approach similar to the preceding section, it is seen that the ratio of the amplitudes (8) take up time dependent values, hence one cannot assume the amplitudes to be of the form (9). If the system is under an external driving force given as
\begin{align}
F_{i,j} = (A_{i,j}\hat{x} + B_{i,j}\hat{y})e^{-i\omega_{0}t}
\end{align}
where $A_{i,j}$ and $B_{i,j}$ are the amplitudes of the external force. The damping force is taken to be a velocity dependent force to first order, given as
\begin{align}
{F_{damp}}_{i,j} = d_{i,j}\frac{dx}{dt}\hat{x} + f_{i,j}\frac{dy}{dt}\hat{y}
\end{align}
$d_{i,j}$ and $f_{i,j}$ are the damping coefficients. One can revisit (1) and (3) to incorporate the given damping and driving force so as to write the equations of motion of the oscillator $(i,j)$ as
\begin{align}
m\frac{d^2x_{i,j}}{dt^2} + d_{i,j}\frac{dx_{i,j}}{dt} + {\omega^2_{0}}(2x'_{i,j} - x'_{i-1,j} - x'_{i+1,j}) = A_{i,j}e^{-i\omega_{0}t}\\
m\frac{d^2y_{i,j}}{dt^2} + f_{i,j}\frac{dy_{i,j}}{dt} + {\omega^2_{0}}(2y'_{i,j} - y'_{i-1,j} - y'_{i+1,j}) = B_{i,j}e^{-i\omega_{0}t}  
\end{align}
For a particular value of $j = j_{0}$, (22) can be represented as  
\begin{align}
{\textbf{K}\cdot{\textbf{X}}} + \textbf{D}{\cdot{\dot{\textbf{X}}}} + \textbf{M}\cdot{\ddot{\textbf{X}}} = \textbf{F}{e^{-i\omega_{0}t}}
\end{align}
Where $K,M,D$ $\&$ $F$ are the mass, dissipation, spring constant and driving force matrices and $\omega_{0}$ is the frequency of the driving force. The matrix representations are given as
\begin{align}
\mathbf{X} = \left(\begin{array}{ccc} x_{0,j} \\ x_{1,j}  \\ x_{2,j} \\ \vdots \\ x_{M-1,j}\end{array} \right), \hspace{0.4cm}
\mathbf{D} = \left(\begin{array}{cccccc} d_{0,j} & 0 & 0 & 0 & \ldots \\ 0 & d_{1,j} & 0 & 0 &\ldots \\ 0 & 0 & d_{2,j} & 0 & \ldots \\ \vdots & \vdots  & \vdots & \vdots & \ddots \end{array} \right) \nonumber
\end{align}
\begin{align}
\mathbf{F} = \left(\begin{array}{ccc} A_{0,j} \\ A_{1,j} \\ A_{2,j} \\ \vdots \\ A_{M-1,j} \end{array} \right), \hspace{0.1cm}
\hspace{0.1cm} \mathbf{K} = \left(\begin{array}{cccccc} {\omega^2}_{0} & -{\omega^2}_{0} & 0 & 0 & 0 &\ldots \\ -{\omega^2}_{0} & 2{\omega^2}_{0} & -{\omega^2}_{0} & 0 & 0 & \ldots \\ 0 & -{\omega^2}_{0} & 2{\omega^2}_{0} & -{\omega^2}_{0} & 0 &\dots \\ \vdots & \vdots & \vdots & \vdots & \vdots & \ddots \end{array}\right) \nonumber
\end{align}
For the sake of simplicity, as mentioned before all the oscillators are assumed to be of the same mass, $m$ hence the mass matrix is an $M\times{N}$ diagonal matrix with the masses of the oscillators as its diagonal entries. The solution of the above equation is of the form \cite{2}
\begin{align}
\mathbf{X} = (\mathbf{K} - i\omega_{0}\mathbf{D} - \omega^2_{0}\mathbf{M})^{-1}\cdot{\mathbf{F}}e^{-i\omega_{0}t} + \mathbf{X}_{free}
\end{align}
where $\mathbf{X}_{free}$ is the solution for homogeneous conditions imposed on (24), viz., under the absence of external driving force.
\begin{align}
\mathbf{X}_{free}= \sum{C_n}e^{-i\omega_nt}\mathbf{X}_n
\end{align}
Hence the problem practically reduces to finding the coefficients $C_n$ in terms of the initial boundary conditions and solving the eigenvalue problem for the operator. The coefficients can be expressed in terms of the initial conditions as
\small{\begin{align}
{{\mathbf{X}= \sum{\mathbf{X}_n}\bigg [ \frac{\mathbf{X}_n\cdot{\mathbf{F}}}{\omega_n - \omega_0}e^{-i\omega_0t} + \mathbf{X_n}\cdot{\big[(\omega_n\mathbf{M} + i\mathbf{D})\cdot\mathbf{X}^{free}(0) + i\mathbf{M}\cdot{\dot{\mathbf{X}}^{free}}(0)\big]}e^{-i\omega_{n}t}\bigg]}}
\end{align}}
\normalsize{
Hence the motion along the x axis is completely determined once the initial conditions are known. Similar treatment of the equation of motion in the y direction,(23) gives the solution for the motion along the y axis. One can also extend this approach to 3-dimensions. 
\section{Conclusions}
It is seen that once the motion is decoupled in the individual directions the motion of the system can be expressed as an expansion amongst the Normal modes. I aim to study the non linear dynamics of the system ~\cite{3}\cite{4}\cite{5},\cite{6} to solve for the motion in the presence of non-linear terms, viz., $O(x'^2/a^2)$ and higher. 
\section{Acknowledgements}
I deeply thank Prof. Chandradew Sharma for introducing me to this problem and for his invaluable help and guidance.
}

\begin{thebibliography}{99}
\bibitem{1} A. P. French, Vibrations and Waves (Indian ed.), CBS Publishers, 1987.
\bibitem{2} C. J. Gobel, S. T. Epstein, ``Motion of Damped Oscillators: Expansion in Normal modes'', American Journal of Physics 48(4) 0002-9505(print) (1980).
\bibitem{3} T. K. Caughey, M. E. J. O'Kelly, ``Effect of Damping on the Natural Frequencies of Linear Dynamic Systems'', J. Acoust. Soc. Am. 33(11) 0001-4966(print) (1961).
\bibitem{4} E. Atlee Jackson, ``Nonlinear Coupled Oscillators. I. Perturbation Theory; Ergodic Problem'', J. Math. Phys. 4, 551 1089-7658(online) (1963).
\bibitem{5} L.I. Manevitch, ``The Description of Localized Normal Modes in a Chain of Nonlinear Coupled Oscillators Using Complex Variables'', Non Linear Dynamics 25(1) DOI: 10.1023/A:1012994430793, Springer Netherlands (2001).
\bibitem{6} Robert Van Buskirk, Carson Jeffries, ``Observation of chaotic dynamics of coupled nonlinear oscillators'', Phys. Rev. A 31, 3332Ð3357 (1985).
\end{thebibliography}
\end{document}